\documentclass{bioinfo}
\copyrightyear{2007}
\pubyear{2007}

\usepackage{mathbbol,amssymb,latexsym,amsfonts,amsmath,amsthm}
\usepackage{graphicx}
\usepackage{float}
\usepackage{textcomp}

\newcommand{\ie}{\textit{i.e.}}
\newcommand{\G}{\mathcal{G}}
\newcommand{\C}{\mathcal{C}}
\newcommand{\D}{\mathcal{D}}
\newcommand{\E}{\mathcal{E}}
\DeclareMathOperator*{\argmax}{arg\, max}

\begin{document}
\firstpage{1}

\title{Analysis of a Gibbs sampler method for model based clustering
  of gene expression data}

\author[A. Joshi \textit{et~al}]{Anagha Joshi\,$^{\rm a,b}$, Yves Van
  de Peer\,$^{\rm a, b}$\footnote{Corresponding author,
    E-mail:yves.vandepeer@psb.ugent.be}, Tom Michoel\,$^{\rm a, b}$}

\address{$^{\rm a}$Department of Plant Systems Biology, VIB,
  Technologiepark 927, 9052 Gent, Belgium, $^{\rm b}$Department of
  Molecular Genetics, UGent, Technologiepark 927, 9052 Gent, Belgium}
\maketitle

\begin{abstract}

  \section{Motivation:} Over the last decade, a large variety of
  clustering algorithms have been developed to detect coregulatory
  relationships among genes from microarray gene expression data.
  Model based clustering approaches have emerged as statistically well
  grounded methods, but the properties of these algorithms when
  applied to large-scale data sets are not always well understood.  An
  in-depth analysis can reveal important insights about the
  performance of the algorithm, the expected quality of the output
  clusters, and the possibilities for extracting more relevant
  information out of a particular data set.
             
  \section{Results:} We have extended an existing algorithm for model
  based clustering of genes to simultaneously cluster genes and
  conditions, and used three large compendia of gene expression data
  for \emph{S.~cerevisiae} to analyze its properties. The algorithm
  uses a Bayesian approach and a Gibbs sampling procedure to
  iteratively update the cluster assignment of each gene and
  condition. For large-scale data sets, the posterior distribution is
  strongly peaked on a limited number of equiprobable clusterings.  A
  GO annotation analysis shows that these local maxima are all
  biologically equally significant, and that simultaneously clustering
  genes and conditions performs better than only clustering genes and
  assuming independent conditions.  A collection of distinct
  equivalent clusterings can be summarized as a weighted graph on the
  set of genes, from which we extract fuzzy, overlapping clusters
  using a graph spectral method.  The cores of these fuzzy clusters
  contain tight sets of strongly coexpressed genes, while the overlaps
  exhibit relations between genes showing only partial coexpression.

  \section{Availability:} \textsf{GaneSh}, a Java package for
  coclustering, is available under the terms of the GNU General Public
  License from our website at
  http://bioinformatics.psb.ugent.be/software.

  \section{Contact:} yves.vandepeer@psb.ugent.be

  \section{Supplementary information:} available on our website at\\
  http://bioinformatics.psb.ugent.be/supplementary\_data/anjos/gibbs
\end{abstract}

\section{Introduction}

Since the seminal paper by \citet{pmid9843981}, now almost a decade
ago, clustering forms the basis for extracting comprehensible
information out of large-scale gene expression data sets. Clusters of
coexpressed genes tend to be enriched for specific functional
categories \citep{pmid9843981}, share \textit{cis}-regulatory
sequences in their promoters \citep{pmid10391217}, or form the
building blocks for reconstructing transcription regulatory networks
\citep{segal2003}.

A variety of heuristic clustering methods have been used, such as
hierarchical clustering \citep{pmid9843981}, $k$-means
\citep{pmid10391217}, or self-organizing maps \citep{pmid10077610}.
Although these methods have had an enormous impact, their statistical
properties are generally not well understood and important parameters
such as the number of clusters are not determined automatically.
Therefore, there has been a shift in attention towards model based
clustering approaches in recent years
\citep{pmid11673243,fraley02,pmid12217911,pmid14871871,chinese,dahl2006}.
A model based approach assumes that the data is generated by a mixture
of probability distributions, one for each cluster, and takes
explicitly into account the noisyness of gene expression data. It
allows for a statistical assessment of the resulting clusters and
gives a formal estimate for the expected number of clusters.  To infer
model parameters and cluster assignments, standard statistical
techniques such as Expectation Maximization or Gibbs sampling are used
\citep{liu2002}.

In this paper we use a novel model based clustering method which
builds upon the method recently introduced by \citet{chinese}. We
address two key questions that have remained largely unanswered for
model based clustering methods in general, namely convergence of the
Gibbs sampler for very large data sets, and non-heuristic
reconstruction of gene clusters from the posterior probability
distribution of the statistical model.

In the model used by \cite{chinese}, it is assumed that the expression
levels of genes in one cluster are random samples drawn from a
Gaussian distribution and expression levels of different experimental
conditions are independent.  We have extended this model to allow
dependencies between different conditions in the same cluster.
\citet{pmid14871871} used a multivariate normal distribution to take
into account correlation among experimental conditions.  Our approach
consists of clustering the conditions within each gene cluster,
assuming that the expression levels of the genes in one gene cluster
for the conditions in one condition cluster are drawn from one
Gaussian distribution.  Hence our model is a model for
\emph{coclustering} or \emph{two-way clustering} of genes and
conditions. The same statistical model was also used in our recent
approach to reconstruct transcription regulatory networks
\citep{lemone}. The coclustering is carried out by a Gibbs sampler
which iteratively updates the assignment of each gene, and within each
gene cluster the assignment of each experimental condition, using the
full conditional distributions of the model.

It is known that a Gibbs sampler may have poor mixing properties if
the distribution being approximated is multi-modal and it will then
have a slow convergence rate \citep{liu2002}.  Previous studies of
Gibbs samplers for model based clustering have not reported
convergence difficulties \citep{pmid12217911,pmid14871871,dahl2006}.
In those studies, only data sets with a relatively small number of
genes (upto a few $100$) \citep{pmid12217911,pmid14871871}, or a small
number of experimental conditions (less than $10$) \citep{dahl2006}
were considered, and special sampling techniques such as reverse
annealing \citep{pmid14871871} or merge-split proposals
\citep{dahl2006} were sufficient to generate a well mixing Gibbs
sampler.  We observe that for data sets of increasing size the
correlation between two Gibbs sampler runs as well as the number of
cluster solutions visited in one run after burn-in steadily decreases.
This means that for large-scale data sets, the posterior distribution
is very strongly peaked on multiple local modes. Since the peaks are
so strong, we approximate the posterior distribution by averaging over
multiple runs performed in parallel, each converging quickly to a
single mode. By computing the correlation between different averages
of the same number of runs we are able to show that the number of
distinct modes is relatively small and accurate approximations to the
posterior distribution can be obtained with as few as $10$ modes for
around $6000$ genes.

To identify the final optimal clustering, the traditional approach is
to select out of all the clusterings visited by the Gibbs sampler the
one which maximizes the posterior distribution (maximum a posteriori
(MAP) clustering).  However, we show that for large data sets the
differences in likelihood between the different local maxima are
extremely small and statistically insignificant, such that the MAP
clustering is as good as taking any local maximum at random. A GO
\citep{ashb00} analysis of the different modes shows that also from
the biological point of view any difference between the local modes is
insignificant.  Taking into account the full posterior distribution is
more difficult since different clusterings may have a different number
of clusters and the labeling of clusters is not unique (the label
switching problem \citep{redner84}).  The common solution to this
problem is to consider pairwise probabilities for two genes being
clustered together or not \citep{pmid12217911,pmid14871871,dahl2006}.
A major question that has not yet recieved a final answer is how to
reconstruct gene clusters from these pairwise probabilities.
\cite{pmid12217911} and \cite{pmid14871871} use a heuristic
hierarchical clustering on the pairwise probability matrix to form a
final clustering estimate.  \cite{dahl2006} introduces a least-squares
method, which selects out of all clusterings visited by the Gibbs
sampler the one which minimizes a distance function to the pairwise
probability matrix. In both approaches, the probability matrix is
reduced to a single hard clustering. This necessarily removes
non-transitive relations between genes (such as a low probability for
a pair of genes to be clustered together even though they both have
relatively high probability to be clustered with the same third gene)
which may nevertheless be informative and biologically meaningful.

We propose that the pairwise probability matrix reflects a \emph{soft}
or \emph{fuzzy clustering} of the data, \ie, genes can belong to
multiple clusters with a certain probability.  To extract these fuzzy
clusters from the pairwise probabilities we use a method from pattern
recognition theory \citep{graphspectral}. This method iteratively
computes the largest eigenvalue and corresponding eigenvector of the
probability matrix, constructs a fuzzy cluster with the eigenvector,
and updates the probability matrix by removing from it the weight of
the genes assigned to the last cluster.  By only keeping genes which
belong to one fuzzy cluster with very high probability we obtain tight
clusters which show higher functional coherence compared to standard
clusters. Keeping also genes which belong with lower but still
significant probability to multiple fuzzy clusters, we can tentatively
identify multifunctional genes or relations between genes showing only
partial coexpression. We show that our results are in good agreement
with previous fuzzy clustering approaches to gene expression data
\citep{gaschfuzzy}. We believe that our fuzzy clustering method to
summarize the posterior distribution will be of general interest for
all model based clustering approaches and solves the problems
associated to heuristic clusterings of the pairwise probability
matrix.

All our analyses are performed on three large-scale public compendia
of gene expression data for \textit{S.~cerevisiae}
\citep{spellmandata,gaschdata,hughesdata}.

\begin{methods}
\section{Methods}

\subsection*{Mathematical model}

For an expression matrix with $N$ genes and $M$ conditions, we define
a coclustering as a partition of the genes into $K$ gene clusters
$\G_k$, together with for each gene cluster, a partition of the set of
conditions into $L_k$ condition clusters $\E_{k,l}$.  We assume that
all data points in a cocluster $\{(i,m)\colon i\in\G_k, m\in
\E_{k,l}\}$ are random samples from the same normal distribution. This
model generalizes the model used by \cite{chinese}, where the
partition of conditions is always fixed at the trivial partition into
singleton sets.

Given a set of means and precisions $(\mu_{kl},\tau_{kl})$, a
coclustering $\C$ defines a probability density on data matrices
$\D=(x_{im})$ by
\begin{align*}
  p\bigl(\D\mid\C,(\mu_{kl},\tau_{kl})\bigr) = \prod_{k=1}^K
  \prod_{l=1}^{L_k} \prod_{i\in\G_k}\prod_{m\in \E_{k,l}} p
  (x_{im}\mid \mu_{kl},\tau_{kl}).
\end{align*}
We use a uniform prior on the set of coclusterings with normal-gamma
conjugate priors for the parameters $\mu_{kl}$ and $\tau_{kl}$.  Using
Bayes' rule we find the probability of a coclustering $\C$ with
parameters $(\mu_{kl},\tau_{kl})$ given the data $\D$.  Then we take
the marginal probability over the parameters $(\mu_{kl},\tau_{kl})$ to
obtain the final probability of a coclustering $\C$ given the data
$\D$, upto a normalization constant:
\begin{equation}\label{eq:1}
  p(\C\mid\D) \propto \prod_{k=1}^K \prod_{l=1}^{L_k} \iint 
  p(\mu,\tau) \prod_{i\in\G_k}\prod_{m\in \E_{k,l}} p (x_{im}\mid
  \mu,\tau)\; d\mu d\tau,
\end{equation}
where $p(\mu,\tau)=p(\mu\mid\tau)p(\tau)$ with
\begin{align*}
  p(\mu\mid\tau)=\bigl(\frac{\lambda_0\tau}{2\pi}\bigr)^{1/2}
  e^{-\frac{\lambda_0\tau}2 (\mu-\mu_0)^2},\quad
  p(\tau) = \frac{\beta_0^{\alpha_0}}{\Gamma(\alpha_0)}
  \tau^{\alpha_0-1} e^{-\beta_0\tau},
\end{align*}
$\alpha_0,\beta_0,\lambda_0 > 0$ and $-\infty<\mu_0<\infty$ being the
parameters of the normal-gamma prior distribution.  We use the values
$\alpha_0=\beta_0=\lambda_0= 0.1$ and $\mu_0=0.0$, resulting in a
non-informative prior. We have compared the normal-gamma prior with
other non-informative, conjugate priors, but found no difference in
results (see Supplementary Information).  The double integral in eq.
(\ref{eq:1}) can be solved exactly in terms of the sufficient
statistics $T^{(n)}_{kl} = \sum_{i \in \G_k,m\in\E_{kl}} x_{im}^n$
($n=0,1,2$) for each cocluster.  The log-likelihood or Bayesian score
decomposes as a sum of cocluster scores:
\begin{equation}\label{eq:7}
  S(\C) =\log p(\C\mid\D) = \sum_{k=1}^K \sum_{l=1}^{L_k} S_{kl},
\end{equation}
with
\begin{multline*}
  S_{kl} = -\tfrac12 T^{(0)}_{kl}\log(2\pi) + \tfrac12
  \log\bigl(\frac{\lambda_0}{\lambda_0 + T^{(0)}_{kl}}\bigr) 
   - \log\Gamma(\alpha_0)\\ + \log\Gamma(\alpha_0
  + \tfrac12 T^{(0)}_{kl})
  + \alpha_0\log\beta_0 -(\alpha_0 + \tfrac12 T^{(0)}_{kl})\log\beta_1
\end{multline*}
and
\begin{equation*}
  \beta_1 = \beta_0 + \frac12\Bigl[ T^{(2)}_{kl} -
  \frac{(T^{(1)}_{kl})^2}{T^{(0)}_{kl}} \Bigr]
  + \frac{\lambda_0 \bigl( T^{(1)}_{kl} - \mu_0 T^{(0)}_{kl}
    \bigr)^2}{2(\lambda_0 + T^{(0)}_{kl})T^{(0)}_{kl}}.
\end{equation*}

\subsection*{Gibbs sampler algorithm}

We use a Gibbs sampler to sample coclusterings from the posterior
distribution (\ref{eq:1}). The algorithm iteratively updates the
assignment of genes to gene clusters, and for each gene cluster, the
assignment of conditions to condition clusters as follows:

\begin{enumerate}
\item Initialization: randomly assign $N$ genes to a random $K_0$
  number of gene clusters, and for each cluster, randomly assign $M$
  conditions to a random $L_{k,0}$ number of condition clusters.
\item For $N$ cycles, remove a random gene $i$ from its current
  cluster.  For each gene cluster $k$, calculate the Bayesian score
  $S(\C_{i\to k})$, where $\C_{i\to k}$ denotes the coclustering
  obtained from $\C$ by assigning gene $i$ to cluster $k$, keeping all
  other assignments of genes and conditions equal, as well as the
  probability $S(\C_{i\to 0})$ for the gene to be alone in its own
  cluster.  Assign gene $i$ to one of the possible $K+1$ gene
  clusters, where $K$ is the current number of gene clusters,
  according to the probabilities $Q_k \propto e^{S(\C_{i\to k})}$,
  normalized such that $\sum_{k} Q_k=1$.
\item For each gene cluster $k$, for $M$ cycles, remove a random
  condition $m$ from its current cluster. For each condition cluster
  $l$, calculate the Bayesian score $S(\C_{k,m\to l})$. Assign
  condition $m$ to one of the possible $L_k+1$ clusters, where $L_k$
  is the current number of condition clusters for gene cluster $k$,
  according to the probabilities $Q_l \propto e^{S(\C_{k,m\to l})}$,
  normalized such that $\sum_{l} Q_l=1$.
\item Iterate step 2 and 3 until convergence. One iteration is defined
  as executing step 2 and 3 consecutively once, and hence consists of
  $N+K\times M$ sampling steps (with $K$ the number of gene clusters
  after Step 1 of that iteration).
\end{enumerate}

This coclustering algorithm simulates a Markov chain which satisfies
detailed balance with respect to the posterior distribution
(\ref{eq:1}), \ie, after a sufficient number of iterations, the
probability to visit a particular coclustering $\C$ is given exactly
by $p(\C\mid\D)$. The expectation value of any real function $f$ with
respect to the posterior distribution can be approximated by averaging
over the iterations of a sufficiently long Gibbs sampler run:
\begin{equation}\label{eq:2}
  E(f) = \sum_\C f(\C) p(\C\mid\D) \approx \frac1T \sum_{t=T_0+1}^{T_0+T}
  f(\C_t)
\end{equation}
where $\C_t$ is the coclustering visited at iteration $t$ and $T_0$ is
a possible burn-in period.  We say that the Gibbs sampler has
converged if two runs starting from different random initializations
return the same averages (\ref{eq:2}) for a suitable set of test
functions $f$. More precisely, if $\{f_n\}$ is a set of test
functions, define $a_n=E_1(f_n)$ the average of $f_n$ in the first
Gibbs sampler run, and $b_n=E_2(f_n)$ the average of $f_n$ in the
second Gibbs sampler run. We define a correlation measure $\rho$
($0\leq\rho\leq1$) between two runs as
\begin{equation}\label{eq:5}
  \rho = \frac{|\sum_n a_n b_n|}{\sqrt{(\sum_n a_n^2) (\sum_n b_n^2)}}.
\end{equation}
Full convergence is reached if $\rho=1$.

\subsection*{Fuzzy clustering}

To keep track of the gene clusters, independent of the (varying)
number of clusters or their labeling, we consider functions
\begin{equation}\label{eq:3}
  f_{ij}(\C) =
  \begin{cases}
    1 & \text{if gene $i$ and $j$ belong to the same gene cluster in $\C$}\\
    0 & \text{otherwise}
  \end{cases}
\end{equation}
In general, the posterior distribution (\ref{eq:1}) is not
concentrated on a single coclustering and the matrix $F=(E(f_{ij}))$
of expectation values (see eq. (\ref{eq:2})) consists of probabilities
between $0$ and $1$. To quantify this fuzzyness, we use an entropy
measure
\begin{equation}\label{eq:4}
  H_{\text{fuzzy}} = \frac1{N^2\ln 2}\sum_{ij }h(F_{ij}),
\end{equation}
where $N$ is the dimension of the square matrix $F$ and
\begin{equation*}
  h(q)=-q\ln(q) - (1-q)\ln(1-q) \text{ for } 0\leq q\leq 1.
\end{equation*}
For a hard clustering ($F_{ij}=0$ or $1$ for all $i,j$),
$H_{\text{fuzzy}}=0$, and for a maximally fuzzy clustering
($F_{ij}=0.5$ for all $i,j$), $H_{\text{fuzzy}}=1$. In reality, the
matrix $F$ is very sparse (most gene pairs will never be clustered
together), so $H_{\text{fuzzy}}$ remains small even for real fuzzy
clusterings.

We assume that a fuzzy gene-gene matrix $F$ is produced by a fuzzy
clustering of the genes, \ie, we assume that each gene $i$ has a
probability $p_{ik}$ to belong to each cluster $k$, such that $\sum_k
p_{ik}=1$. To extract these probabilities from $F$ we use a graph
spectral method \citep{graphspectral}, originally developed for
pattern recognition and image analysis, modified here to enforce the
normalization conditions on $p_{ik}$. A fuzzy cluster is represented
by a column vector $w=(w_1, \dots, w_N)^T$, with $w_i$ the weight of
gene $i$ in this cluster, normalized such that $\|w\|^2=w^Tw=\sum_i
w_i^2=1$.  The cohesiveness of the cluster with respect to the
gene-gene matrix $F$ is defined as $w^TFw = \sum_{ij}w_i F_{ij} w_j$.
By the Rayleigh-Ritz theorem,
\begin{align*}
  \max_{w\neq0} \frac{w^T F w}{w^Tw} = v_1^T F v_1 = \lambda_1,
\end{align*}
where $\lambda_1$ is the largest eigenvalue of $F$ and $v_1$ the
corresponding (normalized) eigenvector. Hence the maximally cohesive
cluster in $F$ is given by the eigenvector of the largest eigenvalue.
By the Perron-Frobenius theorem, this eigenvector is unique and all
its entries are nonnegative. We can then define the membership
probabilities to cluster $1$ by $p_{i1} =
\frac{v_{1,i}}{\max_j(v_{1,j})}$. Hence the gene with the highest
weight in $v_1$ is considered the prototypical gene for this cluster,
and it will not belong to any other cluster. The probability $p_{i1}$
measures to what extent other genes are coexpressed with this
prototypical gene.  To find the next most cohesive cluster, we remove
from $F$ the information already contained in the first cluster by
setting
\begin{align*}
  F^{(2)}_{ij}=\sqrt{1-p_{i1}} F_{ij} \sqrt{1-p_{j1}},
\end{align*}
and compute the largest eigenvalue and corresponding (normalized)
eigenvector $v_2$ for this matrix. The prototypical gene for this
cluster may already have some probability assigned to the previous
cluster, so we define the membership probabilities to the second
cluster by
\begin{align*}
  p_{i2} = \min\Bigl( \frac{v_{2,i}}{\max_j(v_{2,j})}
  (1-p_{i_{\text{max}}1}), 1-p_{i1}\Bigr).
\end{align*}
Here $i_{\text{max}}=\argmax_j(v_{2,j})$ is the prototypical gene for
the second cluster, and we take the `$\min$' to ensure that $\sum_k
p_{ik}$ will never exceed $1$.  

This procedure of reducing $F$ and computing the largest eigenvalue
and corresponding eigenvector to define the next cluster membership
probabilities is iterated until one of the following stopping criteria
is met:
\begin{enumerate}
\item All entries in the reduced matrix $F^{(k)}$ reach $0$, \ie, for
  all genes, $\sum_{k'<k} p_{ik'}=1$, and we have completely
  determined all fuzzy clusters and their membership probabilities.
\item The largest eigenvalue of the reduced matrix $F^{(k)}$ has rank
  $>1$. In this case the eigenvector is no longer unique and need no
  longer have nonnegative entries, so we cannot make new cluster
  membership probabilities out of it. This may happen if the
  (weighted) graph defined by connecting gene pairs with non-zero
  entries in $F^{(k)}$ is no longer strongly connected
  (Perron-Frobenius theorem).
\end{enumerate}

To compute one or more of the largest eigenvalues and eigenvectors for
large sparse matrices such as $F$ and its reductions $F^{(k)}$ we use
efficient sparse matrix routines, such as for instance implemented in
the Matlab$^{\text{\textregistered}}$ function \texttt{eigs}.

\subsection*{Data sets}

We use three large compendia of gene expression data for budding
yeast:
\begin{enumerate}
\item \citet{gaschdata} data set: expression in $173$ stress related
  conditions.
\item \citet{hughesdata} data set: compendium of expression profiles
  corresponding to $300$ diverse mutations and chemical treatments.
\item \citet{spellmandata} data set: $77$ conditions for alpha factor
  arrest, elutriation, and arrest of a cdc15 temperature-sensitive
  mutant.
\end{enumerate}
We select the genes present in all three data sets ($6052$ genes) and,
to be as unbiased as possible, no further postprocessing is done.  We
use SynTReN \citep{syntren} to generate simulated data sets with
varying number of conditions for a synthetic transcription regulatory
network with $1000$ genes (see also Supplementary Information).

\subsection*{Functional coherence}

To estimate the overall biological relevance of the clusters we use a
method which calculates the mutual information between clusters and GO
attributes \citep{clusterjudge}.  For each GOslim attribute, we create
a cluster-attribute contingency table where rows are clusters and
columns are attribute status (\emph{`Yes'} if the gene possesses the
attribute, \emph{`No'} if it is not known whether the gene possesses
the attribute).  The total mutual information is defined as the sum of
mutual informations between clusters and individual GO attributes:
\begin{equation}\label{eq:6}
  MI= \sum_A H(\C)+H(A)-H(\C,A) 
\end{equation}
where $\C$ is a clustering of the genes, $A$ is a GO attribute and $H$
is Shannon's entropy, $H=-\sum_i p_i\log(p_i)$, and the $p_i$ are
probabilities obtained from the contingency tables.

\end{methods}

\section{Results and discussion}

\subsection*{Convergence of the Gibbs sampler algorithm}

We study convergence using the test functions $f_{ij}$ which indicate
if gene $i$ and $j$ are clustered together or not (see eq.
(\ref{eq:3}) in the Methods) and compute the correlation measure
$\rho$ between different runs for this set of functions (see eq.
(\ref{eq:5}) in the Methods).  In addition to the correlation
measure, we also compute the entropy measure $H_{\text{fuzzy}}$
(see eq. (\ref{eq:4}) in the Methods). This parameter summarizes the
`shape' of the posterior distribution: a value of $0$ corresponds to
hard clustering which implies that the distribution is completely
supported on a single solution, the more positive $H_{\text{fuzzy}}$
is, the more the distribution is supported on multiple solutions.

In the analysis below we use subsets from the \citeauthor{gaschdata}
data set with a varying number of genes and conditions and perform
multiple Gibbs sampler runs with a large number of iterations.  One
iteration involves a reassignment of all genes and all conditions in
all clusters, and hence involves $N + M\times K$ sampling steps in the
Gibbs sampler, where $N$ is the number of genes, $M$ the number of
conditions, and $K$ the number of clusters at that iteration
(typically $K\sim\sqrt{N}$).

\begin{figure}[h]
  \centering
  \includegraphics[width=\linewidth]{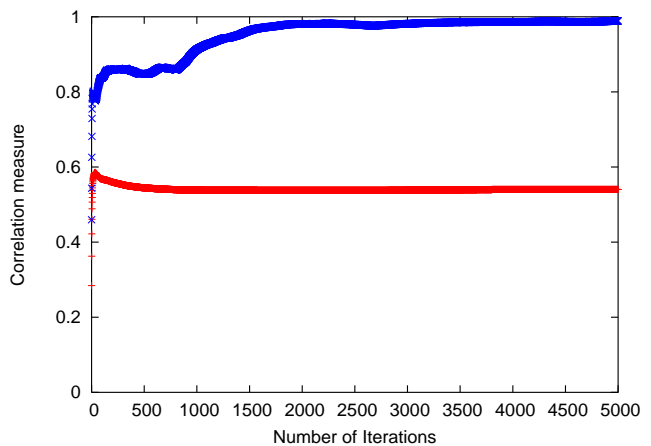}
  \caption{Trace plot of the correlation measure $\rho$ between two
    different Gibbs sampler runs as a function of the number of
    iterations, for a small data set ($100$ genes, $10$ conditions,
    top curve) and a large data set ($1000$ genes, $173$ conditions,
    bottom curve).  Both data sets are subsets of the
    \citeauthor{gaschdata} data set.}
  \label{convergence}
\end{figure}

First we consider a very small data set ($100$ genes, $10$
conditions). We start two Gibbs sampler runs in parallel and compute
the correlation measure $\rho$ at each iteration, see Figure
\ref{convergence}. In this case, $\rho$ approaches its maximum value
$\rho=1$ in less than $5000$ iterations and the Gibbs sampler
generates a well mixing chain which can easily explore the whole
space. Non-zero values of the entropy measure $H_{\text{fuzzy}}$
($0.105\pm0.003$) indicate that the posterior distribution is
supported on multiple clusterings of the genes.

\begin{figure}[h]
\centering
\includegraphics[width=\linewidth]{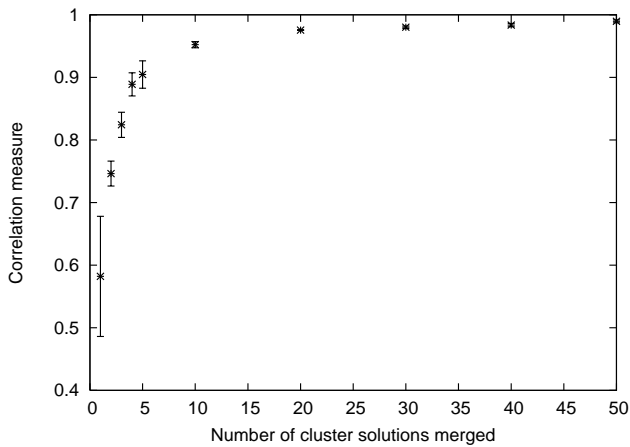}
\caption{Correlation measure $\rho$ between different averages of
  the same number of local maxima for a data set of 1000 genes and 173
  conditions (subset of the \citeauthor{gaschdata} data set).}
\label{merge}
\end{figure}

Next we run the Gibbs sampler algorithm on a data set with $1000$
genes and all 173 conditions.  Unlike in the previous situation we
observe that the correlation between two Gibbs sampler runs saturates
well below $1$ (see Figure \ref{convergence}). Hence the Gibbs sampler
does not converge to the posterior distribution in one run.  We can
gain further understanding for the lack of convergence by looking in
more detail at a single Gibbs sampler run.  It turns out that the
correlation measure between two successive iterations reaches $1$ very
rapidly and remains unchanged afterwards (See Supplementary Figure
$2$).  Since each iteration involves a large number of sampling steps
(\ie, a large number of possible configuration changes), this implies
that the Gibbs sampler very rapidly finds a local maximum of the
posterior distribution from which it can no longer escape.  We
conclude that the posterior distribution is supported on multiple
local maxima which overlap only partially, and with valleys in between
that cannot be crossed by the Gibbs sampler.  These local maxima all
have approximately the same log-likelihood (see for instance the small
variance in Figure \ref{Spellman_conv} below) and are therefore all
equally meaningful.  The probability ratio between peaks and valleys
is so large (exponential in the size of the data set) that an accurate
approximation to the posterior distribution is given by averaging over
the local maxima only. Those can be uncovered by performing multiple
independent runs, each converging very quickly on one of the maxima,
and there is no need for special techniques to also sample in between
local maxima.  The number of local maxima (Gibbs sampler runs)
necessary for a good approximation can be estimated as follows. We
perform $150$ independent Gibbs sampler runs and compute for each the
pairwise gene-gene clustering probability matrix $F$ (see Methods).
For each $k=1,\dots,50$, we take two non-overlapping sets of $k$
solutions and compute the average of their pairwise probability
matrices $F$.  Then, we compute the correlation measure $\rho$ between
those two averages.  This is repeated several times, depending on the
number of non-overlapping sets that can be chosen from the pool of
$150$ solutions.  If for a given $k$ the correlation is always $1$,
then there are at most $k$ local maxima.  Figure \ref{merge} shows
that as $k$ increases, the correlation quickly reaches close to this
perfect value $1$. This implies that the number of local maxima is not
too large and a good approximation to the posterior distribution can
be obtained in this case already with $10$ to $20$ solutions.
Supplementary Figure $1$ shows an example of hard clusters formed as a
result of a single run and fuzzy clusters formed by merging the result
of $10$ independent runs.

\begin{figure}[h]
  \centering
  \includegraphics[width=\linewidth]{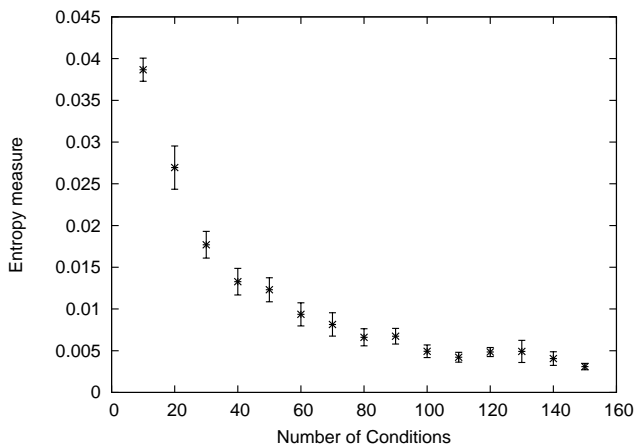}
  \caption{Entropy measure $H_{\text{fuzzy}}$ for data sets with 1000
    genes and varying number of conditions (subsets of the
    \citeauthor{gaschdata} data set).}
  \label{corr_entropy}
\end{figure}

In Figure \ref{corr_entropy}, we keep the same $1000$ genes and select
an increasing number of conditions. As the data set increases, the
entropy measure $H_{\text{fuzzy}}$ decreases, meaning the clusters
become increasingly hard. Simultaneously, the correlation measure
$\rho$ decreases from about $0.85$ to $0.55$ (see Supplementary Figure
$3$).  We conclude that the depth of the valleys between different
local maxima of the posterior distribution increases with the size of
the data set and it becomes increasingly more difficult for the Gibbs
sampler to escape from these maxima and visit the whole space in one
run.

\subsection*{Analysis of whole genome data sets}

\begin{figure}[h]
  \centering
  \includegraphics[width=\linewidth]{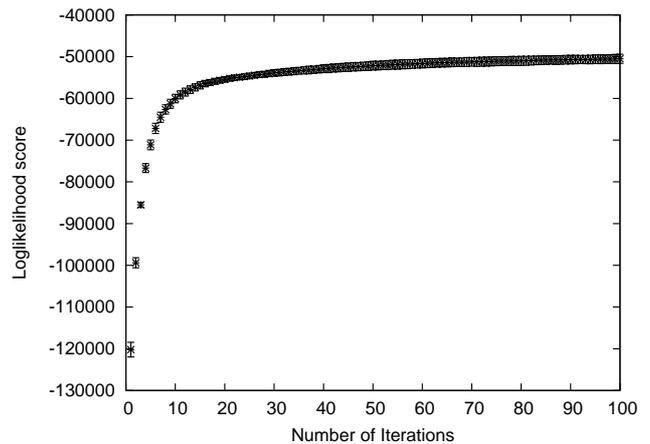}
  \caption{Trace plot of the average log-likelihood score and standard
    deviation for $10$ Gibbs sampler runs for the
    \citeauthor{spellmandata} data set.}
  \label{Spellman_conv}
\end{figure}

If we run the Gibbs sampler algorithm on the three whole genome yeast
data sets, we are in the situation where the algorithm very rapidly
gets stuck in a local maximum. In Figure \ref{Spellman_conv} we plot
the average Bayesian log-likelihood score (see eq. (\ref{eq:7}) in the
Methods) for $10$ different Gibbs sampler runs for the
\citeauthor{spellmandata} data set. The rapid convergence of the
log-likelihood shows that the Gibbs sampler reaches the local maxima
very quickly and the low variance shows that the different local
maxima are all equally likely.  The average over $10$ runs of the GO
mutual information score (see eq.  (\ref{eq:6}) in the Methods) shows
the same rapid convergence and small variance (see Supplementary
Figure $6$), implying that the different maxima are biologically
equally meaningful according to this score. The correlation between
different averages of $10$ Gibbs sampler runs reaches $0.85$, a value
we consider high enough for a good approximation of the posterior
distribution.  The other two data sets show precisely the same
behavior (see Supplementary Figures $4$ and $5$).

\subsection*{Two-way clustering \textit{versus} one-way clustering}

Our coclustering algorithm extends the CRC algorithm of \cite{chinese}
by also clustering the conditions for each cluster of genes
(\emph{`two-way clustering'}), instead of assuming they are always
independent (\emph{`one-way clustering'}). We compare the clustering
of genes for the three yeast data sets using both methods, by
computing the average number of clusters inferred ($K$), the average
log-likelihood score and the average GO mutual information score for
$10$ independent runs of each algorithm.  The results are tabulated in
Table \ref{oneway} and \ref{twoway}.  For all three data sets, both
the log-likelihood score and the GO mutual information score are
higher (better) for our method. The increase in GO mutual information
score is especially significant in case of the \citeauthor{hughesdata}
data set.  This data set has very few overexpressed or repressed
values and if each condition is considered independent, there are very
few distinct profiles which results in the formation of very few
clusters ($\sim 15$ for $6052$ genes). Also clustering the conditions
gives more meaningful results since differentially expressed
conditions form separate clusters from one large background cluster of
non-differentially expressed conditions.

\begin{table}[t]
  \processtable{One-way clustering, averages for $10$ different 
    Gibbs sampler runs.\label{oneway}}
  {\begin{tabular}{lccc}\toprule
      Data set & Avg. $K$ & Avg. log-likelihood score & Avg. MI\\\midrule
      \citeauthor{gaschdata} & $52.9 (2.6)$ & $-6.101 (0.014) \times 10^{5}$ 
      & $1.771 (0.031)$\\
      \citeauthor {hughesdata} & $14.9 (0.5)$ & $2.530 (0.002) \times 10^6$ 
      & $0.588 (0.044)$\\
      \citeauthor{spellmandata} & $49.7 (2.2)$ & $-7.183 (0.037) \times 10^{4}$ 
      & $1.491 (0.032)$\\\botrule
\end{tabular}}{}
\end{table}

\begin{table}[t]
  \processtable{Two-way clustering, averages for $10$ different 
    Gibbs sampler runs.\label{twoway}}
  {\begin{tabular}{lccc}\toprule
      Data set & Avg. $K$ & Avg. log-likelihood score & Avg. MI\\\midrule
      \citeauthor{gaschdata} & $84.5(2.5)$ & $-5.586(0.012)\times 10^{5}$ 
      & $1.912(0.033)$\\
      \citeauthor {hughesdata} & $85.5(2.7)$ & $2.798(0.004)\times 10^6$ 
      & $1.511(0.045)$\\
      \citeauthor{spellmandata} & $65.4(4.2)$ & $-5.112(0.011)\times 10^{4}$ 
      & $1.612(0.032)$\\\botrule
\end{tabular}}{}
\end{table}

\begin{figure}[h]
  \centering
  \includegraphics[width=\linewidth]{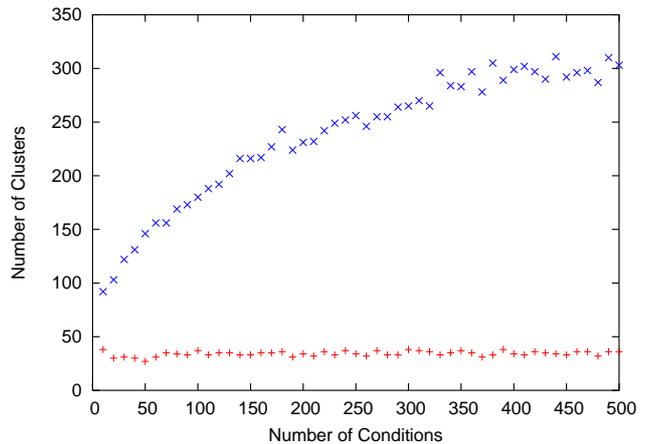}
  \caption{Number of gene clusters for a simulated data set with
    $1000$ genes and a varying number of conditions, for two-way
    clustering (top data points ($\times$)) and one-way clustering
    (bottom data points ($+$))}
  \label{clusterOnewayTwoway}
\end{figure}

For simulated data sets, clusters are defined as sets of genes sharing
the same regulators in the synthetic regulatory network, and the true
number of clusters is known.  Here we consider a gene network whose
topology is subsampled from an \emph{E.~coli} transcriptional network
\citep{syntren} with $1000$ genes, of which $105$ transcription
factors, and $286$ clusters.  For two-way clustering, as we increase
the number of conditions in the simulated data set, more clusters are
formed and the number of clusters saturates close to the true number
(see Figure \ref{clusterOnewayTwoway}). For one-way clustering,
addition of conditions does not affect the inferred number of clusters
which is an order of magnitude smaller than the true number (see
Figure \ref{clusterOnewayTwoway}). For two-way clustering, due to the
clustering of conditions, the number of model parameters is reduced,
and greater statistical accuracy can be achieved, even when the number
of genes in a cluster becomes small.  

The correlation measure $\rho$ between true clusters and inferred
clusters also shows a higher value for two-way clustering over one-way
(Supplementary Figure 8).

Unlike for simulated data sets, the inferred number of clusters does
not depend much upon the number of conditions for real biological data
sets (Supplementary Figure $7$), \ie, even if more conditions are
added, the algorithm does not generate more clusters. This is because
in simulated data, every addition of a condition adds new information,
but for real data sets that might not be the case. In order to get the
true clusters from the expression data, we do not only need more
conditions but also that each new condition contributes information
different from the information already available from the previous
conditions. This might be a reason why the algorithm clusters $6052$
genes in only $\sim 80$ clusters (see Table \ref{twoway}).

\subsection*{Fuzzy clusters}

Our algorithm returns a summary of the posterior distribution in the
form of a gene-gene matrix whose entries are the probabilities that a
pair of genes is clustered together.  To convert these pairwise
probabilities back to clusters we use a graph spectral method as
explained in the Methods. The method produces fuzzy overlapping
clusters where each gene $i$ belongs to each fuzzy cluster $k$ with a
probability $p_{ik}$, such that $\sum_k p_{ik}=1$.  The size of a
fuzzy cluster $k$ is defined as $\sum_i p_{ik}$. The algorithm
iteratively produces new fuzzy clusters until all the information in
the pairwise matrix is converted into clusters ($1^{\text{st}}$
stopping criterium, see Methods), or until the mathematical conditions
underlying the algorithm cease to hold ($2^{\text{nd}}$ stopping
criterium, see Methods). We applied the algorithm to pairwise
probability matrices for each of the three data sets, obtained by
averaging over $10$ different Gibbs sampler runs.  For the
\citeauthor{gaschdata} and \citeauthor{hughesdata} data sets, full
fuzzy clustering is achieved with $500$ fuzzy clusters (all $6052$
genes have total assignment probability $\sum_k p_{ik}>0.98$).  For
the \citeauthor{spellmandata} data set the second stopping
criterium is met after producing $321$ fuzzy clusters.

In general, we observe that the algorithm first produces one very
large fuzzy cluster corresponding to an average expression profile
that almost all genes can relate to. This cluster is of no interest
for further analysis.  Then it produces a number of fuzzy clusters of
varying size which show interesting coexpression profiles and are
useful for further analysis. For the three data sets considered here,
this number is around $100$, consistent with the average number of
clusters in different Gibbs sampler runs (see Table \ref{twoway}). The
remaining fuzzy clusters are typically very small and consist mostly
of noise. Like the very first cluster, they are of no interest for
further analysis.

Since every gene belongs to every cluster, we use a probability cutoff
to remove from each cluster the genes which belong to it with a very
small probability. The smaller the cutoff, the more genes belong to a
cluster, which results into more fuzzy clusters and \textit{vice
  versa}.  Table \ref{cutoff} shows the total number of genes assigned
to at least one fuzzy cluster with different cutoff values and in
brackets the number of genes assigned to at least two fuzzy clusters.

The goal of merging different Gibbs sampler solutions and forming
fuzzy clusters is to extract additional information out of a data set
that is not captured by a single hard clustering solution. This can be
achieved in two ways. First, by obtaining tight clusters of few but
highly coexpressed genes with a high probability cutoff. Second, by
characterizing genes which belong to multiple clusters with a
significant probability.

\begin{table}[!t]
  \processtable{Number of genes clustered and number of genes belonging to 
    multiple clusters with different membership probability cutoff values.\label{cutoff}}
  {\begin{tabular}{lccc}\toprule
      Data set & $0.1$ &  $0.3$  & $0.5$\\ \midrule
      \citeauthor{gaschdata} & $6045$ $(4356)$  &  $4062$ $(344)$  &  $1781$ $(0)$\\
      \citeauthor{hughesdata} & $6052$ $(4554)$  & $3959$ $(34)$  &  $2254$ $(0)$\\
      \citeauthor{spellmandata} & $6052$ $(5187)$  & $3158$ $(139)$  & $1255$ $(0)$\\\botrule
\end{tabular}}{}
\end{table}

\begin{figure}[h]
\centering
\includegraphics[width=\linewidth]{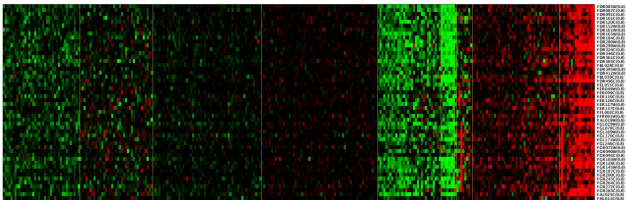}
\caption{Ribosomal genes form a tight cluster in the
  \citeauthor{hughesdata} data set. (Due to space constraints only the
  first few genes are shown; for the complete figure, see the
  Supplementary Information.)}
\label{hughes_ribosome}
\end{figure}

For all three data sets, at a probability cutoff of $0.5$, we get a
subset of genes which belong to only one cluster with high
probability. Table \ref{cutoff} shows that each data set retains at
least $20\%$ of its genes. These are sets of strongly coexpressed
genes which cluster together in almost every hard cluster solution.
Ribosomal genes show such a strong coexpression pattern in all the
three data sets where most genes belong to this cluster with a
probability close to $1$ (see Figure \ref{hughes_ribosome}). At least
$75\%$ of all the genes in cluster $2$ (\citeauthor{gaschdata} data),
cluster $3$ (\citeauthor{hughesdata} data) and cluster $2$
(\citeauthor{spellmandata} data) are located in ribosome.

\begin{figure}[h]
\centering
\includegraphics[width=\linewidth]{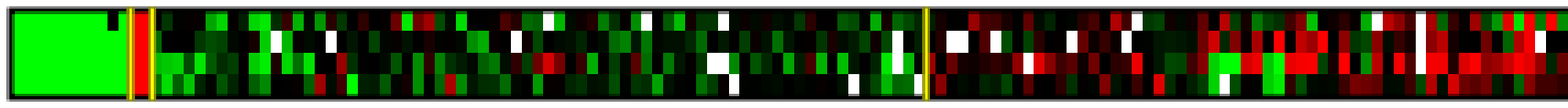}
\caption{Four genes GAL1, GAL2, GAL7 and GAL10 form a tight cluster
  showing conditional coexpression in the \citeauthor{gaschdata} data set.}
\label{gasch_galactose}
\end{figure}

Local but very strong coexpression patterns can also be detected by
our method. Cluster $15$ of the \citeauthor{gaschdata} dataset
consists of only $4$ genes clustered together with probability $1$
(see Figure \ref{gasch_galactose}). These four genes, GAL1, GAL2,
GAL7, and GAL10, are enzymes in the galactose catabolic pathway and
respond to different carbon sources during steady state. They are
strongly upregulated when galactose is used as a carbon source
($2^{\text{nd}}$ experiment cluster in Figure \ref{gasch_galactose})
and strongly downregulated with any other sugar as a carbon source
($1^{\text{st}}$ experiment cluster in Figure \ref{gasch_galactose}).
In every
hard cluster solution, these $4$ genes are clustered together along
with other genes.  By merging these hard cluster solutions to form
fuzzy clusters, we get a tight but more meaningful cluster with only
$4$ genes.

Table \ref{cutoff} shows that many genes belong to two or more
clusters with a significant probability.  For the
\citeauthor{gaschdata} data set, we find similar observations as in
\citep{gaschfuzzy}. Cluster 27 contains genes localized in endoplasmic
reticulum (ER) and induced under dithiothreitol (DTT) stress like
FKB2, JEM1, ERD2, ERP1, ERP2, RET2, RET3, SEC13, SEC21, SEC24 and
others.  Cluster 34 contains genes repressed under nitrogen stress and
stationary state.  20 percent of the genes in cluster 27 also belong
to cluster 34 with a significant membership.  These include genes
encoding for ER vesicle coat proteins like RET2, RET3, SEC13 and
others which are induced under DTT stress as well as repressed under
nitrogen stress and stationary state.  Also RIO1, an essential serine
kinase, belongs to two clusters with a significant probability.  It
clusters with genes involved in ribosomal biogenesis and assembly
(\citeauthor{gaschdata} data cluster $3$) as well as with genes
functioning as generators of precursor metabolites and energy
(\citeauthor{gaschdata} data cluster $7$). We find similar
observations for the \citeauthor{hughesdata} and
\citeauthor{spellmandata} datasets. Genes CLN1, CLN2 and other DNA
synthesis genes like CLB6 which are known to be regulated by SBF
during S1 phase \citep{cellcycle} belong to cluster $19$
(\citeauthor{spellmandata} data).  They also belong with significant
probability to cluster $4$ (\citeauthor{spellmandata} data). More than
one third of the genes in cluster $4$ are predicted to be cell cycle
regulated genes.

\section*{Conclusion}

We have developed an algorithm to simultaneously cluster genes and
conditions and sample such coclusterings from a Bayesian probabilistic
model.  For large data sets, the model is supported on multiple
equivalent local maxima. The average of these local maxima can be
represented by a matrix of pairwise gene-gene clustering probabilities
and we have introduced a new method for extracting fuzzy, overlapping
clusters from this matrix. This method is able to extract information
out of the data set that is not available from a single, hard
clustering.

\section*{Funding}

Early Stage Marie Curie Fellowship to A.J.; Postdoctoral Fellowship of
the Research Foundation Flanders (Belgium) to T.M.

\section*{Acknowledgement}
We thank Steven Maere and Vanessa Vermeirssen for helpful discussions.


\end{document}